\documentclass[10pt,pre,showpacs,byrevtex,nobibnotes,nofootinbib]{revtex4}
\usepackage{graphics}
\usepackage{bm} 
\usepackage{amssymb}
  
\newcommand{\be}{\begin{equation}}
\newcommand{\ee}{\end{equation}}
\newcommand{\eff}{{ }}
\newcommand{\pr}{\partial}
\newcommand{\tile}{\tilde \vare}
\newcommand{\tilm}{\tilde m}
\newcommand{\tils}{\tilde s}
\newcommand{\vare}{\varepsilon}
\newcommand{\tilz}{\tilde Z}
\newcommand{\tilvp}{\tilde\varphi}
\newcommand{\bh}{\bar h}
\newcommand{\ra}{\rightarrow}

\begin{document}

\title{Negative magnetic susceptibility and nonequivalent ensembles for the\\ 
mean-field $\boldsymbol{\phi^4}$ spin model} 

\author{Alessandro Campa}
\email{alessandro.campa@iss.infn.it}
\affiliation{Complex Systems and Theoretical Physics Unit, Health and Technology
Department, Istituto Superiore di Sanit\`a and INFN Roma,
Gruppo Collegato Sanit\`a, Viale Regina Elena 299, 00161 Roma, Italy}

\author{Stefano Ruffo}
\email{stefano.ruffo@unifi.it}
\affiliation{Dipartimento di Energetica ``Sergio Stecco'' and CSDC, Universit\`a di
Firenze and INFN,
Via S. Marta 3, 50139 Firenze, Italy}

\author{Hugo Touchette}
\email{ht@maths.qmul.ac.uk}
\affiliation{School of Mathematical Sciences, Queen Mary, University of London,
London E1 4NS, UK}

\date{\today}

\begin{abstract}
We calculate the thermodynamic entropy of the mean-field $\phi^4$ spin model in the microcanonical
ensemble as a function of the energy and magnetization of the model. The entropy and its derivative
are obtained from the theory of large deviations, as well as from Rugh's microcanonical formalism,
which is implemented by computing averages of suitable observables in microcanonical molecular
dynamics simulations. Our main finding is that the entropy is a concave function of the energy for
all values of the magnetization, but is nonconcave as a function of the magnetization for some values
of the energy. This last property implies that the magnetic susceptibility of the model can be negative
when calculated microcanonically for fixed values of the energy and magnetization. This provides a
magnetization analog of negative heat capacities, which are well-known to be associated in general
with the nonequivalence of the microcanonical and canonical ensembles. Here, the two ensembles that
are nonequivalent are the microcanonical ensemble in which the energy and magnetization are held fixed
and the canonical ensemble in which the energy and magnetization are fixed only on average by fixing
the temperature and magnetic field.
\end{abstract}

\pacs{05.20.-y, 05.20.Gg, 05.70.Fh}
\maketitle 

\section{Introduction}

The Legendre transform connecting the entropy function of the microcanonical ensemble and the free
energy of the canonical ensemble in the thermodynamic limit of these ensembles is often used in practice
to obtain the entropy of a system from the knowledge of its free energy. Unfortunately, as has been
stressed repeatedly in studies of systems involving long-range interactions \cite{gross1997,gross2001,
eyink1993,ellis2000,dauxois2002,touchette2004}, the entropy is the Legendre transform of the free energy
only if the entropy is concave as a function of the energy. If the entropy is a nonconcave function of
the energy, as it often happens in long-range systems, then it cannot be calculated as the Legendre
transform of the free energy. In this case, one must resort to analytically obtain the entropy by other
means, e.g., by evaluating directly the density of states from which the entropy is defined (see, e.g.,
Refs.~\cite{ispolatov2000,barre2001}), or by using large deviation techniques based on the microcanonical
ensemble \cite{ellis2000,ellis2004,costeniuc2005,barre2005,hahn2006}. Another possibility put forward
recently works by modifying the definition of the canonical ensemble in such a way that nonconcave
entropies can be obtained from the Legendre transform of a modified form of free energy
\cite{costeniuc22005,costeniuc2006}. Examples of applications of this generalized canonical ensemble
can be in found in Refs.~\cite{touchette2006,costeniuc22006}.

There are many types of statistical models whose entropy is known to be a nonconcave function of the
energy. Examples, listed in the order in which they were discovered, include systems of particles
interacting through gravitational forces \cite{lynden1968,thirring1970,chavanis2002,chavanis22002},
a model of plasma \cite{smith1990}, statistical models of two-dimensional turbulence
\cite{kiessling1997,ellis2002}, as well as several spin models involving long-range and mean-field
interactions \cite{ispolatov2000,barre2001,ellis2004,costeniuc2005}. Our goal in this paper is to
study yet another model, which departs somewhat from the models listed above in that its microcanonical
entropy is nonconcave as a function of its energy \emph{and} magnetization. This, as we shall see, has
many consequences for the equivalence of the many different ensembles that can be conceived for this model
according to whether the energy and magnetization are treated in a canonical or microcanonical way,
i.e., whether these quantities are assumed to fluctuate or not. Parallels between these cases of
nonequivalent ensembles and those studied in the context of the energy alone will be discussed. In
particular, we shall see that one consequence of the nonconcavity of the entropy with the added
magnetization is that the magnetic susceptibility of the model, calculated microcanonically, can be
negative. The parallel for systems having nonconcave
entropies as a function of the energy is that the heat capacity can be negative
\cite{lynden1968,thirring1970,gross1997,gross2001,lynden1999,touchette22003}. 

The possibility for the entropy to be nonconcave in a variable other than the energy or in more than
one variable is discussed by Ellis, Haven and Turkington \cite{ellis2000} in the context of their
general theory of nonconcave entropies and nonequivalent ensembles. These authors presented in
Ref.~\cite{ellis2002} a first application of their theory for a statistical model of two-dimensional
turbulence, whose entropy is nonconcave as a function of the energy and circulation. A model similar
to the one treated here having a nonconcave entropy as a function of its energy and magnetization
was also studied recently by Hahn and Kastner in Refs.~\cite{hahn2005,hahn2006}. What we present 
here can be seen as a continuation and an extension of these papers. The difference with 
Refs.~\cite{hahn2005,hahn2006} is that we present not one but two methods for calculating the 
entropy as a function of the energy and magnetization: one based on large deviation theory and a
second based on Rugh's microcanonical formalism \cite{rugh1998,rugh2001}. Additionally, we discuss some
of the relationships that exist between the nonconcavity of the entropy, the nonequivalence of the
microcanonical and canonical ensembles, and the appearance of first-order phase transitions in the
canonical ensemble. 

\section{The model and its thermodynamic description}
\label{smodel}

The model that we study in this paper is the so-called mean-field $\phi^4$ model, defined by the
Hamiltonian
\be
\label{ham}
H = \sum_{i=1}^N \left(\frac{p_i^2}{2}-\frac{1}{4}q_i^2 + \frac{1}{4}q_i^4 \right) 
-\frac{1}{4N} \sum_{i,j=1}^N q_i q_j.
\ee
In this expression, $q_i$ and $p_i$ are the canonical coordinates of unit mass particles moving on
a line ($q_i,p_i \in \Bbb{R}$). These particles are subjected to a local double-well potential, in
addition to interact with each other through a mean-field (infinite range) interaction given by
the all-to-all coupling in the double sum. In the following, we shall use $x$ to denote a point of
the phase space $\Gamma$ of the system, i.e., $x=(\{p_i\},\{q_i\})$.

The mean-field $\phi^4$ model was introduced by Desai and Zwanzig \cite{desai1978} who studied its
relaxation to equilibrium using Langevin dynamics. More recently, Dauxois, Lepri and Ruffo
\cite{dauxois2003} have studied this model in the canonical ensemble, and showed that it exhibits
a second-order ferromagnetic phase transition. The critical temperature of the transition is found
to be $T_c \approx 0.264$, corresponding to a critical energy per particle or mean energy
$\vare_c = T_c/2 \approx 0.132$. The steps leading to the calculation of the entropy of the
mean-field $\phi^4$ model augmented by an extra magnetic field were also presented recently by two
of us in Ref.~\cite{campa2006}. In the same paper, molecular dynamics simulations performed above
the critical energy are reported in an attempt to study the convergence of finite-$N$ averages
calculated in Rugh's microcanonical formalism. Finally, as mentioned in the introduction, Hahn and
Kastner \cite{hahn2005,hahn2006} have studied a mean-field $\phi^4$ model similar to the one
studied here. They have calculated for their model the thermodynamic entropy as a function of the
 mean energy $\vare(x)=H(x)/N$ and the mean magnetization, defined as
\be\label{mmag}
m(x)=\frac{1}{N}\sum_{i=1}^N q_i.
\ee
We recall that, in terms of these two quantities, the definition of the entropy is
\be
s(\vare,m)=\lim_{N\ra\infty}\frac{1}{N}\ln\int_\Gamma \delta(\vare(x)-\vare)\delta(m(x)-m)dx.
\ee
Naturally, in terms of $\vare$ alone, we have
\be
s(\vare) = \lim_{N\ra\infty}\frac{1}{N}\ln\int_\Gamma \delta \left(\vare(x)-\vare\right)dx.
\label{ent1}
\ee

Our goal in the next sections is to calculate $s(\vare,m)$ for the Hamiltonian defined in (\ref{ham}).
In doing so, we shall try to highlight the difficulties that arise in calculating this function due to the
fact that it is nonconcave, in addition to discuss the physical consequences of having a nonconcave entropy
in two thermodynamic variables. The fundamental difficulty, for what concerns calculations, is basically the following. 
If $s(\vare,m)$ were concave, then we could calculate this function from the point of view of the canonical ensemble
using the following steps:

(i) Calculate the partition function
\be
\label{partfct1}
Z(\beta,\eta)=\int_\Gamma e^{-\beta H(x) -\eta M(x)}dx, 
\ee
where $M(x)=Nm(x)$ is the total magnetization. 

(ii) Calculate the thermodynamic free energy function of the model defined by
\be
\varphi(\beta,\eta)=-\lim_{N\ra\infty} \frac{1}{N}\ln Z(\beta,\eta).
\ee

(iii) Obtain $s(\vare,m)$ by taking the Legendre transform of free energy function $\varphi(\beta,\eta)$; in symbols, 
\be
s(\vare,m)=\beta \vare + \eta m -\varphi(\beta,\eta),
\label{lf1}
\ee
with $\beta$ and $\eta$ determined by the equations
\be
\frac{\pr}{\pr \beta}\varphi(\beta,\eta)=\vare,\quad  \frac{\pr}{\pr \eta}\varphi(\beta,\eta)=m.
\label{lf2}
\ee

The problem with nonconcave entropies is that the last step does not actually yield the correct entropy
function because Legendre transforms yield only concave functions. To circumvent this problem, we proceed
in the next section to obtain $s(\vare,m)$ using another method suggested by large deviations
\cite{ellis2000}. The method is the same as that used in Refs.~\cite{hahn2005,hahn2006,campa2006}
(see also Refs.~\cite{ellis2004,costeniuc2005,barre2005}). The results obtained will then be compared with
those derived with Rugh's method. This second method is the subject of Sec.~\ref{sproc}. The results
obtained by the two methods are reported, compared and discussed in Sec.~\ref{scomp}.

Before we jump to the next section, it is useful to note that the partition function shown in (\ref{partfct1})
can be re-written in the more familiar form
\be
Z(\beta,h)=\int_\Gamma e^{-\beta [H(x) -h M(x)]}dx
\label{canz1}
\ee
by defining $\eta=-\beta h$. This shows that the partition function defined in (\ref{partfct1}) is nothing
but the standard, canonical partition function of $H$ with an added external magnetic field $h$. As is well
known, $h$ is the field of the canonical ensemble which is conjugated to the magnetization constraint
$m(x)=m$ of the microcanonical ensemble, while $\beta$ is the canonical field conjugated to the microcanonical
energy constraint $\vare(x)=\vare$. These observations are important for what is coming later. 

\section{Large deviation calculation of the entropy}
\label{sldt}

The large deviation method described in this section is essentially a generalization of the maximum entropy principle,
which is particularly suited for many-particle systems involving long-range or mean-field
interactions. The method is presented in detail in Ref.~\cite{ellis2000}, and was used recently to calculate
the entropy function of many models; see, e.g., Refs.~\cite{ellis2004,costeniuc2005,barre2005}.

The basic ingredients of the method are the following. First, one must be able to find a set of macro-variables
or ``mean fields'', denoted collectively by the vector $\mu(x)$, which are such that the energy per particle
$\vare(x)$ and the mean magnetization $m(x)$ can be re-written as a function of these variables. In symbols,
this means that there must exist two functions\footnote{Functions referring to the macrostate $\mu$ bear a
tilde to distinguish them from those referring to $\vare$ and $m$.} 
$\tile$ and $\tilm$ of $\mu$ such that
\be
\vare(x)=\tile(\mu(x)),\quad m(x)=\tilm(\mu(x)).
\ee
(See Ref.~\cite{ellis2000} or \cite{ellis2004} for a more complete and more accurate statement of this
condition.) Second, one must be able to derive the expression of the entropy function $\tils (\mu)$ for the macrostate
$\mu$, which, in analogy with $s(\vare)$ and $s(\vare,m)$, is defined as
\be
\tils (\mu)=\lim_{N\ra\infty} \frac{1}{N}\ln \int_\Gamma \delta(\mu(x)-\mu) dx.
\ee
In large deviation theory, the function $\tils (\mu)$ is interpreted as the rate function (up to a sign)
governing the fluctuations of $\mu$ with respect to the probability measure defining the microcanonical
ensemble \cite{ellis2000}. With this function, we finally obtain the entropy $s(\vare, m)$ by solving a
constrained maximization problem given by
\be
s(\vare,m)=\sup_{\mu: \tile (\mu)=\vare,\tilm (\mu)=m} \tils (\mu).
\label{mep1}
\ee
This formula is the generalized maximum-entropy principle that we alluded to above. In large deviation
theory, such a formula is referred to as a contraction formula or contraction principle
\cite{ellis1985,ellis1999}.

The real challenge in solving the variational problem of (\ref{mep1}) is not so much to solve the constrained maximization,
but to derive a priori the expression of $\tils (\mu)$. An implicit hope of the large
deviation method, in this respect, is that $\tils (\mu)$ be a concave function of the selected macrostate
$\mu$. In this case, the calculation of $\tils (\mu)$ is facilitated by the fact that $\tils (\mu)$ is
the Legendre transform of some properly-defined free energy function. Specifically, define
\be
\tilz (\lambda)=\int_\Gamma e^{-N \lambda\cdot\mu(x)} dx 
\ee
to be the partition function of the observable $\mu(x)$. In this expression, $\lambda\cdot\mu(x)$ stands
for the usual scalar product of the two vectors $\lambda$ and $\mu(x)$. Now, let
\be
\tilvp (\lambda)=-\lim_{N\ra\infty}\frac{1}{N}\ln \tilz(\lambda)
\ee
be the free energy function associated with $\mu(x)$. Then, assuming that $\tils (\mu)$ is concave, we have
\be
\tils (\mu) = \inf_{\lambda} \{\lambda\cdot\mu -\tilvp(\lambda) \}.
\label{lf3}
\ee
This equation is the macrostate generalization of the Legendre transform defined by Eqs.~(\ref{lf1}) and
(\ref{lf2}). It is now a valid equation for calculating $\tils (\mu)$ because the latter is assumed to
be concave. 

All of the steps just described work for the mean-field $\phi^4$ model. To start, it is easy to see that
a good choice of macrostate for this model is the vector $\mu=(m,k,v)$ composed of the mean magnetization
$m$, the mean kinetic energy
\be
k=\frac{1}{2N}\sum_{i=1}^N p_i^2,
\ee
and the mean potential energy
\be
v=\frac{1}{4N}\sum_{i=1}^N (q_i^4-q_i^2). 
\ee
In terms of $\mu$, we indeed have
\be
\tile (\mu) = k +v -\frac{m^2}{4},
\label{efct1}
\ee
and since $\mu$ already includes the mean magnetization $m$, there is no need to define a function $\tilm$.

To calculate the entropy $\tils (\mu)$, we follow the Legendre transform path, anticipating that
$\tils (\mu)$ is concave. Since each component of $\mu$ is additive, the partition function for
$\mu$ has the form
\be
\tilz(\lambda)=\tilz(\lambda_m,\lambda_k,\lambda_v)=[\tilz_k(\lambda_k)]^N [\tilz_{m,v}
(\lambda_m,\lambda_v)]^N,
\ee
where
\be
\tilz_k(\lambda_k) =\int_{-\infty}^{\infty} e^{-\lambda_k p^2/2}dp=\sqrt{\frac{2\pi}{\lambda_k}}
\ee
and
\be
\tilz_{m,v}(\lambda_m,\lambda_v)=\int_{-\infty}^\infty e^{-\lambda_m q -\lambda_v (q^4-q^2)/4} dq.
\label{quartint1}
\ee
Note that in order for $\tilz_k(\lambda_k)$ to exist, $\lambda_k>0$; similarly, $\lambda_v>0$ above. From
the expression of the partition function, we write the expression of the free energy function of $\mu$ as 
\be
\tilvp(\lambda_m,\lambda_k,\lambda_v)=\tilvp_k(\lambda_k)+\tilvp_{m,v}(\lambda_m,\lambda_v),
\ee
where 
\be
\tilvp_k(\lambda_k)=-\ln\tilz_k(\lambda_k)=\frac{1}{2}\ln\lambda_k-\frac{1}{2}\ln(2\pi),
\ee
and $\tilvp_{m,v}(\lambda_m,\lambda_v)=- \ln\tilz_{m,v}(\lambda_m,\lambda_v)$. In order to be able to
express $\tils (m,k,v)$ as the Legendre transform of $\tilvp(\lambda_m,\lambda_k,\lambda_v)$, we now have
to verify that $\tils (m,k,v)$ is concave. One way of verifying this is to verify that $\tilvp_k(\lambda_k)$ is a 
differentiable function of all its arguments.\footnote{This result is at the root of the problem of ensemble 
equivalence. If the free energy is differentiable, then its Legendre transform yields the correct concave 
entropy (strictly concave, in fact). See Ref.~\cite{ellis2004,touchette22003} for more information.} This, 
as is easily verified, is indeed the case, so we can proceed to calculate the Legendre transform shown
in Eq.~(\ref{lf3}); the result is
\be
\tils (m,k,v)=\tils_k (k)+ \tils (m,v),
\ee
where
\be
\tils_k (k) = \inf_{\lambda_k} \{ \lambda_k k - \tilvp_k(\lambda_k)\}=\frac{1}{2}\ln k + \frac{1}{2}
\ln (4\pi e)
\label{kent1}
\ee
and
\be
\tils_{m,v}(m,v)=\inf_{\lambda_m\in\Bbb{R},\lambda_v>0} \{ \lambda_m m +\lambda_v v -\tilvp_{m,v}
(\lambda_m,\lambda_v)\}.
\label{mvent1}
\ee
Since $\tilvp_{m,v}(\lambda_m,\lambda_v)$ is differentiable, the last equation can actually be re-written
as
\be
\tils_{m,v}(m,v)=\lambda_m(m,v) m +\lambda_v(m,v) v -\tilvp_{m,v}(\lambda_m(m,v),\lambda_v(m,v)),
\ee
where $\lambda_m(m,v)$ and $\lambda_v(m,v)$ are the unique solutions of the two equations
\be
\frac{\pr}{\pr \lambda_m}\tilvp(\lambda_m,\lambda_v)=m,\quad  \frac{\pr}{\pr \lambda_v}\tilvp
(\lambda_m,\lambda_v)=v.
\ee
We clearly see in these equations the familiar form of the Legendre transform; compare them with
Eqs.~(\ref{lf1}) and (\ref{lf2}).

We have now reached the last step of the calculation of $s(\vare,m)$, which is to solve the constrained
maximization problem displayed in equation (\ref{mep1}). In our case, this equation takes the form
\be
s(\vare,m)=\sup_{ k,v:k+v=\vare+m^2/4} \{ \tils_k(k) +\tils_{m,v}(m,v)\}.
\ee 
With the explicit expression of $\tils_k(k)$ shown in (\ref{kent1}), this can be re-written as
\be
s(\vare,m)=\sup_{v} \left\{ \frac{1}{2}\ln\left( \vare-v+\frac{m^2}{4}\right)+\frac{1}{2}\ln(4\pi e)
+\tils_{m,v}(m,v)\right\}.
\label{finmax1}
\ee
Thus, in the end, to find $s(\vare,m)$ we have to solve a one-dimensional, unconstrained maximization
problem involving the two-dimensional function $\tils_{m,v}(m,v)$. The calculation of $\tils_{m,v}(m,v)$
calls itself for the solution of the two-dimensional minimization problem shown in Eq.~(\ref{mvent1}),
which involves the free energy $\tilvp_{m,v}(\lambda_m,\lambda_v)$. Note that the mean potential energy
$v$ is constrained to lie in the range $[-1/16,\vare+m^2/4)$. The lower bound of this range arises
naturally as the minimum of the double-well potential, whereas the upper bound arises because
$k=\vare-v+m^2/4>0$. Finally, we note that the parity of the Hamiltonian (\ref{ham})
in the coordinates $q$ implies that the function $s(\vare,m)$ will be even in $m$.

\section{Rugh's formalism and molecular dynamics simulations}
\label{sproc}

The results of the large deviation method just outlined will be compared in the next section with results
obtained from Rugh's microcanonical formalism \cite{rugh1998,rugh2001}. The idea behind this formalism is
to perform molecular dynamics simulations of the Hamiltonian system $H$, and to compute the time average
of certain observables along the system's trajectory. Assuming that the dynamics of the system is ergodic,
one then equates the time average of these observables with their microcanonical ensemble averages. In this
context, what Rugh's formalism provides is a general prescription for estimating important thermodynamic
quantities of the microcanonical ensemble as time averages of suitably-chosen dynamical observables. 

In a previous paper \cite{campa2006}, we have described the implementation of Rugh's formalism for the
mean-field $\phi^4$ model augmented by a magnetic field. The main results of this paper, adapted to our
specific model without the magnetic field, are the following. Let $H(x,M)$ denote the Hamiltonian of a
system whose magnetization is constrained to have the value $M$. How this constraint is to be put in the 
original Hamiltonian $H(x)$ will be discussed below. The microcanonical entropy of the model as a function
of the total energy $E$ and total magnetization $M$ is defined as
\be
S(E,M)=\ln\int_\Gamma \delta\left( H(x;M)-E\right) dx,
\ee  
while the average of a general observable $A(x)$ in the microcanonical ensemble is given by
\be
\left\langle A \right\rangle_{E,M} =\frac{\displaystyle\int_\Gamma  \delta
\left( H(x,M)-E\right) A(x)dx}{\displaystyle\int_\Gamma \delta\left( H(x;M)-E\right) dx}.
\label{aveg}
\ee
We consider now the total (extensive) entropy of the system rather than the (intensive) entropy density
$s(\vare,m)$ because we want to keep track of the $N$-dependence of the entropy. As usual, relevant
thermodynamic quantities like the temperature, the specific heat and the magnetic susceptibility are
defined through derivatives of the entropy. In Rugh's formalism, these derivatives are calculated by
choosing a vector $Y$ in $\Gamma$ such that $Y\cdot\nabla H=1$. In terms of $Y$, we then have
\be
\frac{\partial}{\partial E}S(E,M) =
\left\langle\nabla \cdot Y\right\rangle_{E,M} = \frac{1}{T(E,M)}
\label{davega}
\ee
and
\be
\frac{\partial}{\partial M}S(E,M)= -\left\langle\nabla
\cdot \left(\frac{\partial H}{\partial M} Y\right) \right\rangle_{E,M}.
\label{davegb}
\ee
The first derivative with respect to $E$ defines, as always, the inverse temperature $T(E,M)^{-1}$ of the
system. The derivatives of $\langle A\rangle_{E,M}$ are computed similarly as
\be
\frac{\partial}{\partial E}\left\langle A \right\rangle_{E,M}=\left\langle \nabla \cdot
(A Y) \right\rangle_{E,M}-\frac{1}{T(E,M)}\left\langle A \right\rangle_{E,M}
\label{davegc}
\ee
and
\be
\frac{\partial}{\partial M}\left\langle A \right\rangle_{E,M} = -\left\langle \nabla\cdot
\left(\frac{\partial H}{\partial M}A Y\right) \right\rangle_{E,M}+\left\langle \nabla \cdot
\left(\frac{\partial H}{\partial M} Y\right) \right\rangle_{E,M}\left\langle A \right\rangle_{E,M}
+ \left\langle \frac{\partial A}{\partial M} \right\rangle_{E,M}.
\label{davegd}
\ee
Equations (\ref{davega})-(\ref{davegd}) are valid for any number $N$ of particles. As mentioned before,
for ergodic systems, the ensemble averages $\langle \ldots \rangle_{E,M}$ entering in these equations are
computed as time-averages obtained through molecular dynamics simulations in which the energy and the
magnetization are both conserved. The convergence of these time-averages towards their thermodynamic
($N\ra\infty$) values is studied in Ref.~\cite{campa2006}.

Molecular dynamics simulations performed with $H(x)$ automatically conserve the energy, so it remains to adapt
them to make sure that they also conserve the magnetization, i.e., that $Nm(x)=M$ at all time. One way to achieve
this is to explicitly incorporate the magnetization constraint into the Hamiltonian $H(x)$ by eliminating, for example,
the variable $q_N$ using $q_N=M-\sum_{i=1}^{N-1}q_i$. In this way, we obtain a new constrained Hamiltonian $H_c(x,M)$
involving $N-1$ particles and the magnetization $M$:
\be
H_c=\frac{1}{2}\sum_{i=1}^{N-1}p_i^2 -\frac{1}{2N}\sum_{i,j}^{1,N-1}p_ip_j + \sum_{i=1}^{N-1}
\left( -\frac{1}{4}q_i^2 + \frac{1}{4}q_i^4 \right)-\frac{1}{4}\left(M-\sum_{i=1}^{N-1}q_i\right)^2
+\frac{1}{4}\left(M-\sum_{i=1}^{N-1} q_i\right)^4 - \frac{M^2}{4N}.
\label{hamcon}
\ee
Note that the the kinetic energy is no more diagonal because the magnetization constraint couples to the momenta. 

For this Hamiltonian, we follow Rugh \cite{rugh1998}, and choose the vector $Y$ to have nonvanishing components
only in correspondence with the kinetic energy. That is, we choose
\be
Y = \frac{1}{2K_c}\left(p_1,\dots,p_{N-1},0,\dots,0\right),
\label{ycon}
\ee 
where $K_c$ is the kinetic part of $H_c$. It is easy to verify that $Y \cdot\nabla H_c =1$. Consequently,
we can use Eq.~(\ref{davega}) to obtain the temperature $T(E,M)$; the result is
\be
\frac{1}{T(E,M)} = \left\langle \frac{N-3}{2K_c}\right\rangle_{E,M}.
\label{tcon}
\ee
Similarly, we obtain from (\ref{davegb}),
\be
\frac{\partial}{\partial M} S(E,M) = \frac{m}{T(E,M)}-\left\langle m_3 \frac{N-3}{2K_c} \right\rangle_{E,M}
\label{dms1}
\ee
with $T(E,M)$ given by (\ref{tcon}) and $m_3$ defined by 
\be
m_3 = \frac{1}{N}\sum_{i=1}^N q_i^3.
\label{m3def}
\ee
By defining an \textit{effective magnetic field} $h_\eff (E,M)$ with the usual thermodynamic relation
\be
h_\eff (E,M) = -T(E,M) \frac{\partial}{\partial M}S(E,M),
\label{hdef}
\ee
we can also put (\ref{dms1}) in the form
\be
h_\eff (E,M) = -m + T(E,M)\left\langle m_3 \frac{N-3}{2K_c} \right\rangle_{E,M}.
\label{hrugh}
\ee
Further derivatives of this quantity and of $T(E,M)$, as calculated with Eqs.~(\ref{davegc}) and (\ref{davegd}),
yield the magnetic susceptibility and the heat capacity, respectively. 

At this point, it is important to note that $T(E,M)$ and $h_\eff (E,M)$ are microcanonical
quantities---they are functions of the constrained values of the energy and magnetization. These quantities
must be distinguish from their canonical counterparts, $T=(k_B \beta)^{-1}$ and $h$, which are not
functions of any other variables---they are the parameters of the canonical ensemble. The equivalence of
these two ensembles will be explored in the next section.

For now, we finish this section by discussing another method which can be used to implement the magnetization constraint
into the molecular dynamics simulations. The method relies on the use of Lagrange multipliers,
and works by simulating the dynamics of the Hamiltonian
\be
H_\nu=H+\nu M,
\ee 
which is our original Hamiltonian augmented by the magnetization $M$ and its associated Lagrange
multiplier $\nu$. The equations of motion for $H_\nu$ read
\be
\ddot{q}_i = -\frac{\partial H_\nu}{\partial q_i} = -\frac{\partial V}{\partial q_i} - \nu \frac{\partial M}{\partial q_i}=
 -\frac{\partial V}{\partial q_i} - \nu,\quad \quad \quad i=1,\dots,N.
\label{eqgcon}
\ee
In this expression, $V$ denotes the potential energy, i.e., $V=H-K$. An expression for $\nu$ in terms of
the $q_i$'s is readily obtained from (\ref{eqgcon}) by noting that $M$ is constant in time, so that
$
\sum_{i=1}^N \ddot{q_i}=0.
$
As a result,
\be
\nu = -\frac{1}{N}\sum_{i=1}^N \frac{\partial V}{\partial q_i}= \frac{1}{N}\sum_{i=1}^N q_i
-\frac{1}{N}\sum_{i=1}^N q_i^3 =m-m_3,
\label{concomp}
\ee
where, for the last equalities, we have used the explicit expression of the potential energy of the
mean-field $\phi^4$ model. Inserting this back into Eq.~(\ref{eqgcon}) leads us, finally, to the
following equations of motion\footnote{These equations could also be obtained from $H_c$ by
re-inserting the coordinate $q_N$ into the canonical equations of motion.}:
\be
\ddot{q}_i = \frac{1}{2}q_i - q_i^3 - \frac{1}{2}m + m_3, \quad \quad \quad i=1,\dots,N.
\label{eqcon2}
\ee

The main advantage of considering $H_\nu$ for performing the molecular dynamics simulations is that it
allows a direct calculation of the effective magnetic field $h_\eff (E,M)$. Indeed, in view of the
equations of motion derived from $H_\nu$ and the form of $H_\nu$, it is natural to interpret the Lagrange
multiplier $\nu$ as the opposite of a fictitious magnetic field which adjusts itself in time in order to
keep the magnetization constant. The time average of this instantaneous field defines another effective
magnetic field, denoted by $\bh(E,M)$, whose explicit expression is
\be
\bh (E,M) = - \langle \nu \rangle_{E,M}= -m + \left\langle m_3 \right\rangle_{E,M}.
\label{heff}
\ee
Now, although this expression is different from the expression of the field $h_\eff (E,M)$ obtained in
Rugh's formalism, we have observed in our simulations that the difference between the two fields is
negligible for $N$ large, so that $h_\eff (E,M)$ and $\bh (E,M)$ can be considered to be equal for
all practical purposes. It can be noted, in fact, that Eq.~(\ref{hrugh}) reduces to Eq.~(\ref{heff}) if
\be
\left\langle m_3 \frac{N-3}{2K_c}\right\rangle_{E,M} = \left\langle m_3 \right\rangle_{E,M}
\left\langle \frac{N-3}{2K_c}\right\rangle_{E,M}. 
\label{hequal}
\ee
Thus, if $m_3$ becomes statistically uncorrelated with the kinetic observable, which is something
expected to occur in the thermodynamic limit, then $h_\eff(E,M)=\bh (E,M)$.

The calculation of the magnetic susceptibility $\chi(E,M)$ at constant energy and magnetization, defined by
\be
\frac{1}{\chi (E,M)}= N\frac{\partial}{\partial M}h(E,M),
\ee
is also simplified by considering $\bh (E,M)$ instead of $h_\eff (E,M)$. The magnetic susceptibility is
calculated in Rugh's formalism from Eq.~(\ref{davegd}). When applied to $\bh (E,M)$, this equation yields
\be
\frac{1}{\chi (E,M)}= (N-3)\left[ \left\langle \frac{m_3(m-m_3)}{2K_c}\right\rangle_{E,M}
-\left\langle m_3\right\rangle_{E,M} \left\langle \frac{(m-m_3)}{2K_c} \right\rangle_{E,M} \right]
- 1 +3\left\langle m_2 \right\rangle_{E,M},
\label{susccon}
\ee
where, in analogy with $m_3$, we have introduced the notation
\be
m_2 = \frac{1}{N}\sum_{i=1}^N q_i^2.
\label{q2def}
\ee
We leave it to the reader to check that the expression of $\chi(E,M)$ obtained from $h_\eff (E,M)$ is
more complicated. In the end, it is interesting to note that we could have calculated $\chi(E,M)$ by taking
the numerical derivative of $h_\eff (E,M)$ or $\bh(E,M)$. This, however, generally has the effect of
amplifying the error associated with either field.  In Rugh's formalism, $\chi(E,M)$ is calculated as
a time-average just like $h_\eff (E,M)$ or $\bh (E,M)$, and so carries a numerical error comparable
to the error associated with these fields.

\section{Comparison and discussion of the results}
\label{scomp}

\begin{figure*}[t]
\resizebox{0.9\textwidth}{!}{\includegraphics{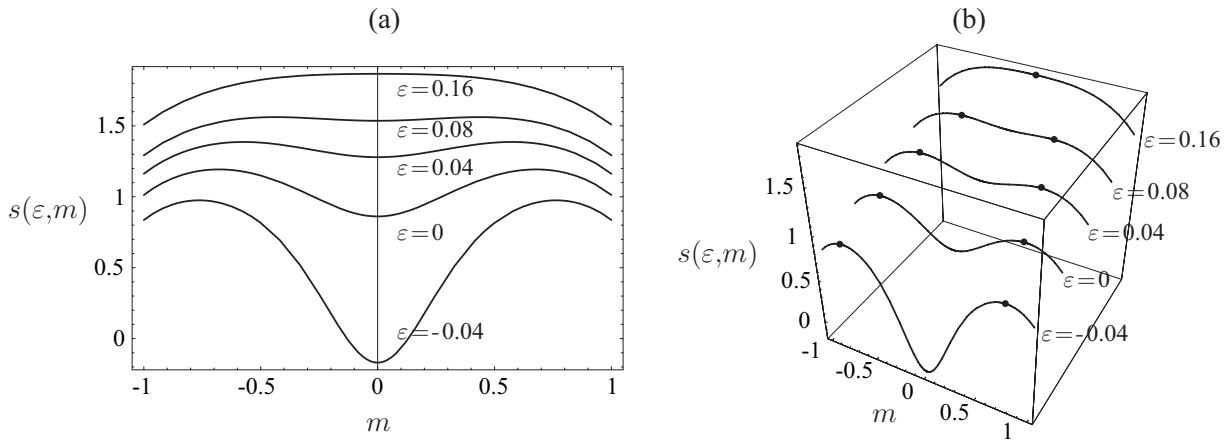}}
\caption{(a) Entropy as a function of the mean magnetization $m$ for different values of the mean energy
$\vare$. (b) 3D view of $s(\vare,m)$. The black dots show the location of the equilibrium values of $m$
in the microcanonical ensemble for each value of $\vare$ (see the text for explanation).}
\label{sem3fig}
\end{figure*}

The entropy density $s(\vare,m)$ of the mean-field $\phi^4$ model as a function of its mean energy $\vare$
and mean magnetization $m$ is shown in Fig.~\ref{sem3fig}. In Fig.~\ref{sem3fig}(a), we have plotted the
graph of $s(\vare,m)$ as a function of $m$ for four different values of $\vare$: one above the critical
value $\vare_c\approx 0.132$ and three below. To get a better idea of the shape of $s(\vare,m)$ as a
two-dimensional function, we also report in Fig.~\ref{sem3fig}(b) the curves of Fig.~\ref{sem3fig}(a)
stacked along the $\vare$ direction. The data shown in both of these plots are those obtained via the
large deviation method, i.e., via the maximization problem displayed in (\ref{finmax1}). It should be
noted that this maximization problem cannot be solved analytically because the free energy
$\tilvp_{m,v}(\lambda_m,\lambda_v)$, which intervenes in the derivation of the macrostate entropy function
$\tils_{m,v} (m,v)$, involves a quartic integral; see Eq.~(\ref{quartint1}). Nevertheless, this integral,
like all the other quantities and optimization problems involved in the calculation of $s(\vare,m)$, can
easily be handled numerically with the end result that $s(\vare,m)$ can easily be computed numerically to
any desired precision. The specific results that we present in Fig.~\ref{sem3fig} were obtained with
Mathematica using the default 16-digit working precision. The resulting numerical error is
smaller than the tickness of the curves reported in Fig.~\ref{sem3fig}.

\begin{figure}[t]
\resizebox{3.2in}{!}{\includegraphics{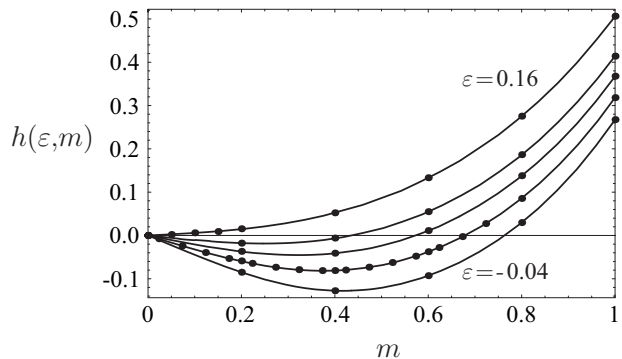}}
\caption{Effective magnetic field calculated using Rugh's method ($\bullet $) and the large deviation
method (full line). The values of $\vare$ are from top to bottom $\vare=0.16,0.08,0.04,0$ and $-0.04$.
In this figure and in the next one the error associated with the results of the molecular dynamics
simulations is smaller than the filled dots.}
\label{hvm1fig}
\end{figure}

\begin{figure}[t]
\resizebox{3.2in}{!}{\includegraphics{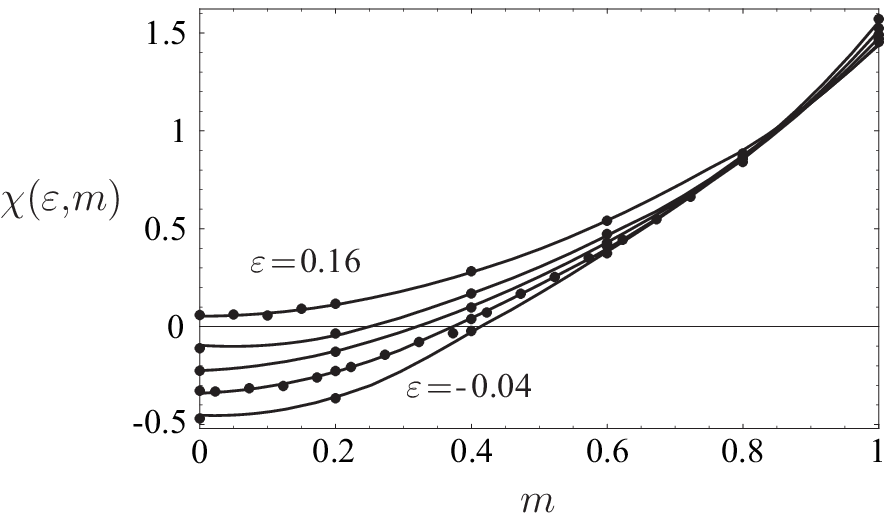}}
\caption{Magnetic susceptibility calculated using Rugh's method ($\bullet $) and the large
deviation method (full line).}
\label{dhvm1fig}
\end{figure}

The comparison of the large deviation method with Rugh's method is reported in Figs.~\ref{hvm1fig} and
\ref{dhvm1fig}. Figure \ref{hvm1fig} shows the results of the calculation of the effective magnetic
field $h(\vare,m)$ using the large deviation method (full line) and Rugh's method (filled dots). Because
$s(\vare,m)$ is an even function of $m$, we show $h(\vare,m)$ only for $m>0$. In the former method,
$h(\vare,m)$ is calculated from its thermodynamic-limit definition
\be\label{htherm}
h(\vare,m)=-T(\vare,m) \frac{\partial s(\vare,m)}{\partial m},
\ee
using the form of $s(\vare,m)$ shown in Fig.~\ref{sem3fig}. In the latter method, the same field is
calculated using Eq.~(\ref{hrugh}) or Eq.~(\ref{heff}). As mentioned earlier, these two equations are
expected to yield comparable results so long as the molecular dynamics simulations are performed for
large-enough systems. In our case, $N=10^2$ particles were used, and we found no appreciable differences
between the fields obtained with Eqs.~(\ref{hrugh}) and (\ref{heff}). From Fig.~\ref{hvm1fig}, we see that there
is a perfect agreement between the field $h(\vare,m)$ obtained from the large deviation approach and the
same field obtained from Rugh's approach. We thus conclude that the two approaches agree
on the calculation of $s(\vare,m)$, up to a constant which can easily be determined. The two approaches
perfectly agree also for the calculation of the magnetic susceptibility at constant energy; see Fig.~\ref{dhvm1fig}. 

Having shown that the large deviation approach for calculating $s(\vare,m)$ gives the same results as those
obtained with Rugh's approach, we now come to the discussion of the nonequivalence of the microcanonical and
canonical ensembles in relation to the nonconcavity of $s(\vare,m)$.\footnote{By microcanonical ensemble, we mean of
course the ensemble in which both the energy \emph{and} magnetization are fixed. The concommittant canonical
ensemble is the one in which $\beta$ and $h$ are fixed.} A first hint to the effect that the two ensembles are
nonequivalent is provided by the fact that the effective field $h(\vare,m)$ can be negative for positive values of
$m$ when $\vare<\vare_c$, as can be seen from Fig.~\ref{hvm1fig}. In the canonical ensemble, it can be
shown that the equilibrium magnetization of the mean-field $\phi^4$ model can only be negative when the
canonically-imposed field $h$ is negative, and can only be positive when $h$ is positive.\footnote{For generic
Hamiltonians that are not even in $q$, the magnetization and magnetic field could have opposite signs. What we report
here is specific to the mean-field $\phi^4$ model, which has the parity in $q$.}
In other words, in the canonical ensemble, the magnetization of the $\phi^4$ model is always in the direction of the field. 
In the microcanonical ensemble, we see that this is not always the case: $h(\vare,m)$ and $m$ can have opposite signs, so
that this ensemble must be nonequivalent with the canonical ensemble.

\begin{figure}[t]
\resizebox{3.3in}{!}{\includegraphics{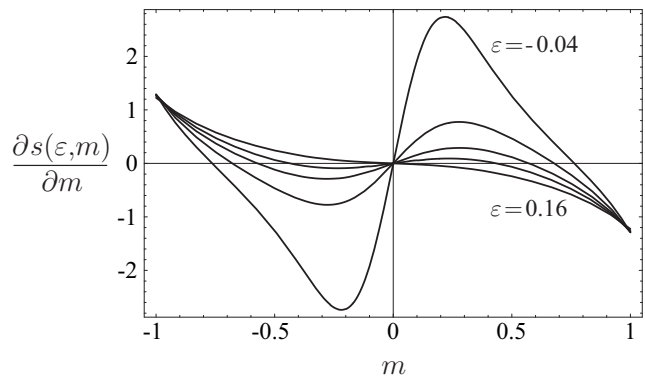}}
\caption{$m$-derivative of the entropy function shown in Fig.~\ref{sem3fig}(a).}
\label{dsem1fig}
\end{figure}

To understand how this peculiarity of the $\phi^4$ model arises out of the
nonconcavity of $s(\vare,m)$, we only have to analyze the definition of $h(\vare,m)$, given by
Eq.~(\ref{htherm}). There $T(\vare,m)$ is always positive, since it is directly proportional to the kinetic
energy, so that the sign of $h(\vare,m)$ is always the opposite of the sign of $\partial s(\vare,m)/\partial m$.
Therefore, in order for $h(\vare,m)$ to be negative for $m>0$, we must have $\partial s(\vare,m)/\partial m >0$
for $m>0$. This, as can be seen by comparing Figs.~\ref{sem3fig}(a), \ref{hvm1fig} and \ref{dsem1fig}, can only
happen if $s(\vare,m)$ is nonconcave, i.e., if the graph of $s(\vare,m)$ has a bimodal or ``double-bump'' shape
as a function of the magnetization.\footnote{A similar argument can be
used to show that $h(\vare,m)$ can be positive for $m<0$ if and only if $s(\vare,m)$ is nonconcave in $m$.}
When $s(\vare,m)$ is concave or unimodal, as is the case for $\vare=0.16$, then $\partial s(\vare,m)/\partial m$
is necessarily negative when $m>0$ (Fig.~\ref{dsem1fig}), which implies that $h(\vare,m)$ is necessarily positive
when $m>0$ (Fig.~\ref{hvm1fig}), in agreement with what is seen in the canonical ensemble. 

Another consequence of the nonconcavity of $s(\vare,m)$ is that the magnetic susceptibility $\chi(\vare,m)$,
calculated microcanonically at fixed values of $\vare$ and $m$, can be negative for certain values of $m$ and
$\vare$; see Fig.~\ref{dhvm1fig}. This, alone, is a sure sign that the microcanonical and canonical ensembles
are nonequivalent, for we know that the magnetic susceptibility is always positive in the canonical ensemble.
Indeed, we know that by increasing the magnetic field $h$ while keeping the inverse temperature $\beta$ constant,
the equilibrium magnetization $m(\beta,h)$ of the canonical ensemble can only increase, which means that 
\be
\chi(\beta,h)=\frac{\partial m(\beta,h)}{\partial h} \geq 0.
\ee
(See the Appendix for a general proof of this result.) In the microcanonical ensemble, the role of $h$ and $m$ are
reversed: what is varied is $m$ not $h$, and the proper magnetic susceptibility to consider in this ensemble, defined as
\be
\chi(\vare,m)=\left(\frac{\partial h(\vare,m)}{\partial m}\right)^{-1}=\left(\frac{\partial s}{\partial \vare}\right)^2\left(
\frac{\partial^2 s}{\partial m \partial \vare}\frac{\partial s}{\partial m}-
\frac{\partial s}{\partial \vare}\frac{\partial^2 s}{\partial m^2}\right)^{-1},
\label{chimic}
\ee 
can be negative, as demonstrated in Fig.~\ref{dhvm1fig}. Physically, this means that, in the microcanonical ensemble,
an increase in $m$ can lead to a decrease of the microcanonical field $h(\vare,m)$. This, as we have already mentioned,
is related to the fact that $s(\vare,m)$ is nonconcave as a function of $m$; however, it is important to note that
$\chi(\vare,m)$ does not become negative \textit{exactly} when $s(\vare,m)$ becomes nonconcave. Indeed, 
the local convexity properties of $s(\vare,m)$ are determined by the eigenvalues of the Hessian matrix 
\be
\left(\begin{array}{cc}
\partial^2 s/\partial m^2 & \partial^2 s/\partial m \partial \vare \\
 \partial^2 s/\partial \vare \partial m & \partial^2 s/\partial \vare^2
\end{array} \right)~,
\label{suscept}
\ee
whose characteristic equation is
\be
x^2-\left[\frac{\partial^2 s}{\partial m^2}+\frac{\partial^2 s}{\partial \vare^2}\right]x
+\left[\frac{\partial^2 s}{\partial m^2}\frac{\partial^2 s}{\partial \vare^2}
-\left(\frac{\partial^2 s}{\partial m \partial \vare}\right)^2\right]=0~.
\label{chara}
\ee
The sign of the eigenvalues of the Hessian changes when either the sign of the first-order term or
of the zeroth-order term of the characteristic equation changes. This occurs on two lines in the $(\vare,m)$
plane, which can either intersect at one point on the line where the susceptibility given in Eq.~(\ref{chimic})
changes sign, or not intersect at all. In either case, the change in the local convexity properties 
of $s(\vare,m)$ and in the sign of $\chi(\vare,m)$ does not happen generically at the same value of
$m$ for a given value of $\vare$. This is also seen by comparing Figs.~\ref{dhvm1fig} and \ref{dsem1fig}.

To close our discussion of the mean-field $\phi^4$ model, we now comment on the thermodynamic properties of
this model obtained from the microcanonical ensemble in which only the energy is fixed. In this ensemble,
the relevant entropy function to consider is the function $s(\vare)$ defined in Eq.~(\ref{ent1}). From the
knowledge of $s(\vare,m)$, we can derive $s(\vare)$ by using the following contraction formula \cite{ellis2000,ellis2004}:
\be
s(\vare)=\sup_{m} s(\vare,m).
\ee
Figure \ref{sem3fig}(b) shows the positions of the maxima of $s(\vare,m)$, which represent physically the
equilibrium values of the magnetization in the microcanonical ensemble with fixed energy $N\vare$. We see
from this figure that $s(\vare,m)$ has only one maximum located at $m=0$ when it is concave as a function of
$m$. This happens when $\vare\geq \vare_c$, so that $s(\vare)=s(\vare,m=0)$ for all $\vare\geq \vare_c$.
Below $\vare_c$, the single maximum of $s(\vare,m)$ splits in a continuous way into two symmetric maxima. The
continuous character of the splitting can be deduced from the fact that the phase transition of the model, seen
in the canonical ensemble having $\beta$ as its only parameter, is known to be of second order
\cite{dauxois2003}. This implies that the phase transition seen in the microcanonical ensemble as a function
of $\vare$ is also second-order, i.e., continuous. This must be so because a second-order phase transition in
the canonical ensemble implies that the canonical and microcanonical ensembles are fully equivalent, since
$s(\vare)$ in this case is concave and has no affine parts, i.e., no straight (linear) lines in its graph
\cite{touchette22003,ellis2004,touchette22006,costeniuc2006}. These two features of $s(\vare)$ are somewhat
difficult to see from Fig.~\ref{sem3fig}; however, they are confirmed by the theory of nonconcave entropies.
The fact indeed is that if $s(\vare)$ was nonconcave as a function of $\vare$, then the phase transition of
the model would be first-order as a function of the temperature rather than second-order
\cite{touchette22003,touchette22006}. The same applies if $s(\vare)$ was affine: in this case, the phase
transition in the canonical ensemble would also be first-order \cite{touchette22003,touchette22006}, which is
not what is reported by Dauxois \textit{et al.}~\cite{dauxois2003}.

The relationship between nonconcave entropies and first-order phase transitions is fully general, and can be
applied to $s(\vare,m)$ to conclude that the mean-field $\phi^4$ model displays, in the canonical ensemble
with temperature $T$ and magnetic field $h$, a first-order, \emph{field-driven} phase transition when
$T<T_c=2\vare_c$. The phase transition, in this case, is driven by the magnetic field $h$ because the entropy
$s(\vare,m)$ is nonconcave as a function of the magnetization, the variable conjugated to $h$. Moreover, since
$h$ is basically the derivative of $s(\vare,m)$, it should be clear from the symmetric, bimodal shape of
$s(\vare,m)$, seen again when $\vare<\vare_c$, that the critical field of the phase transition is $h_c=0$.
Physically, this means that if we set the temperature $T$ to be below $T_c$, then the equilibrium magnetization
$m(T,h)$ of the canonical ensemble jumps from a negative, non-zero value to a positive, non-zero value as
$h$ varies from $0^-$ to $0^+$. The ``spontaneous'' values of $m(T,h)$ found for $h=0^-$ and $h=0^+$ are
nothing but the two values of $m$ at which $s(\vare,m)$ if maximum when $\vare<\vare_c$.

\section{Conclusion}
\label{sconc}

The entropy of long-range systems can be concave as a function of the energy alone, and yet be nonconcave as
a function of the energy and other invariant quantities. This was illustrated here for the mean-field
$\phi^4$ model: its entropy is concave as a function of the energy, but is nonconcave as a function of the
energy and magnetization. This property of the entropy leads, as we have shown, to a fundamental difference
between the thermodynamic properties of the model found in the microcanonical ensemble as a function of the
energy and magnetization, and those found in the canonical ensemble as a function of the temperature and
magnetic field. On the one hand, we have shown that the (effective) magnetic field of the microcanonical
ensemble can have a sign opposite to that of the magnetization. In the canonical ensemble, this never happens,
as the equilibrium magnetization in this ensemble is always in the direction of the applied magnetic field.
On the other hand, we showed that the magnetic susceptibility at constant temperature, calculated
microcanonically as a function of the energy and magnetization, can be negative, whereas it is always positive in
the canonical ensemble. Because of these two differences, we say that the microcanonical and canonical
ensembles are nonequivalent.

There is a further consequence of the nonconcavity of the entropy that we have discussed, namely that the
mean-field $\phi^4$ model displays a first-order phase transition driven by the magnetic field in the canonical
ensemble. This phase transition is directly related to the fact that the entropy of the model is nonconcave as
a function of the magnetization---the quantity conjugated to the magnetic field---for certain values of the
energy. Thus, in addition to display a second-order phase transition driven by the temperature (or the energy),
the mean-field $\phi^4$ model displays a first-order phase transition driven by the magnetic field. The
situation, as such, is similar to the two-dimensional Ising model, which also displays a second-order phase
transition in temperature but a first-order phase transition in the magnetic field. There is, however, a
fundamental difference between the Ising model and the mean-field $\phi^4$ model. Being a short-range
interaction model, the Ising model has an entropy which is necessarily concave as a function of its energy
and magnetization \cite{ruelle1969}. The first-order, field-driven phase transition of this model is thus
not related to a nonconcavity of the entropy; it appears, in fact, because the entropy, when plotted as a
function of the magnetization, has affine parts in the form of ``plateaus.'' For more details on this point,
we refer the reader to Sec.~11 of Ref.~\cite{ellis1995} and the references cited therein.

Our last words of this paper will go to a technical remark about a possible way to compute the entropy
$s(\vare,m)$ using Legendre transforms. We mentioned in Sec.~\ref{smodel} that because $s(\vare,m)$ is
nonconcave, it cannot be calculated as the Legendre transform of the energy function $\varphi(\beta,\eta)$
or, equivalently, $\varphi(\beta,h)$, which is the free energy associated with the partition function shown
in (\ref{canz1}). However, there is a subtle workaround to this problem. Indeed, although $s(\vare,m)$ is
nonconcave as a function of $m$, it is concave in $\vare$ for all $m$. This suggests that the projections of
$s(\vare,m)$ along $\vare$ can be calculated by taking the Legendre transform of a free energy function, which,
unconventionally, is a function of the inverse temperature and the magnetization. The construction of this
Legendre transform, which follow from the theory of Ellis \emph{et al.}~\cite{ellis2000}, is outlined in
Appendix \ref{amixed}.

\begin{acknowledgments}
H.T. would like to thank the occupiers of ``17-04 The Regalia'' in Singapore, and especially Ana Belinda
Pe\~nalver, for having provided a perfect environment in which to work. The hospitality of the Institute for
Mathematical Sciences at the National University of Singapore, where part of this work was completed, is also
acknowledged. Support for this work was provided by NSERC (Canada) and the Royal Society of London.
\end{acknowledgments}

\appendix

\section{Proof of the positivity of $\chi(\beta,h)$}
\label{positive}

The equilibrium magnetization $m(\beta,h)$, calculated in the canonical ensemble as a function of the inverse
temperature $\beta$ and magnetic field $h$, can be deduced from the equation
\be
m(\beta,h)=-\frac{1}{\beta}\frac{\partial \varphi(\beta,h)}{\partial h},
\ee
where $\varphi(\beta,h)$ is the canonical free energy function associated with the partition function shown
in (\ref{canz1}). With this equation, the definition of the magnetic susceptibility $\chi(\beta,h)$
in the canonical ensemble can be rewritten as
\be
\chi(\beta,h)=-\frac{1}{\beta} \frac{\partial^2 \varphi(\beta,h)}{\partial h^2}.
\ee
At this point, we note that $\varphi(\beta,h)$ is an always convex function of $\beta$ and $h$, so that
\be
\frac{\partial^2 \varphi(\beta,h)}{\partial h^2}\leq 0
\ee
for all $\beta$ and $h$. Consequently, $\chi(\beta,h)\geq 0$ for all values of $\beta>0$ and $h$, as
claimed. 

\section{Mixed canonical-microcanonical ensemble}
\label{amixed}

Define the partition function
\be
Z(\beta,m)=\int_\Gamma e^{-\beta H(x)} \delta(m(x)-m) dx
\ee
and its associated free energy
\be
\varphi(\beta,m)=-\lim_{N\ra\infty}\frac{1}{N}\ln Z(\beta,m).
\ee
Then
\be
s(\vare,m)=\beta\vare-\varphi(\beta,m),
\ee
where the value of $\beta$ is set by solving
\be
\frac{\partial \varphi(\beta,m)}{\partial\beta} = \vare. 
\ee
The last two equations define the Legendre transform that takes $\varphi(\beta,m)$ to $s(\vare,m)$. It is a
valid Legendre transform for the mean-field $\phi^4$ model because $s(\vare,m)$ is concave as a function of
$\vare$ for all values of $m$ for this model; see Fig.~\ref{sem3fig}. Following
Ellis \textit{et al.}~\cite{ellis2000}, $Z(\beta,m)$ and $\varphi(\beta,m)$ are to be interpreted,
respectively, as the partition function and free energy function of a \emph{mixed} canonical-microcanonical
ensemble in which the energy is treated in a canonical way, while the magnetization is treated in a
microcanonical way. The reader is referred to Ref.~\cite{ellis2000} for more details on the idea of mixed
ensembles.

\bibliography{phi4}

\begin{thebibliography}{36}
\expandafter\ifx\csname natexlab\endcsname\relax\def\natexlab#1{#1}\fi
\expandafter\ifx\csname bibnamefont\endcsname\relax
  \def\bibnamefont#1{#1}\fi
\expandafter\ifx\csname bibfnamefont\endcsname\relax
  \def\bibfnamefont#1{#1}\fi
\expandafter\ifx\csname citenamefont\endcsname\relax
  \def\citenamefont#1{#1}\fi
\expandafter\ifx\csname url\endcsname\relax
  \def\url#1{\texttt{#1}}\fi
\expandafter\ifx\csname urlprefix\endcsname\relax\def\urlprefix{URL }\fi
\providecommand{\bibinfo}[2]{#2}
\providecommand{\eprint}[2][]{\url{#2}}

\bibitem[{\citenamefont{Gross}(1997)}]{gross1997}
\bibinfo{author}{\bibfnamefont{D.~H.~E.} \bibnamefont{Gross}},
  \bibinfo{journal}{Phys. Rep.} \textbf{\bibinfo{volume}{279}},
  \bibinfo{pages}{119} (\bibinfo{year}{1997}).

\bibitem[{\citenamefont{Gross}(2001)}]{gross2001}
\bibinfo{author}{\bibfnamefont{D.~H.~E.} \bibnamefont{Gross}},
  \emph{\bibinfo{title}{Microcanonical Thermodynamics: Phase Transitions in
  ``Small'' Systems}}, vol.~\bibinfo{volume}{66} of
  \emph{\bibinfo{series}{Lecture Notes in Physics}} (\bibinfo{publisher}{World
  Scientific}, \bibinfo{address}{Singapore}, \bibinfo{year}{2001}).

\bibitem[{\citenamefont{Eyink and Spohn}(1993)}]{eyink1993}
\bibinfo{author}{\bibfnamefont{G.~L.} \bibnamefont{Eyink}} \bibnamefont{and}
  \bibinfo{author}{\bibfnamefont{H.}~\bibnamefont{Spohn}}, \bibinfo{journal}{J.
  Stat. Phys.} \textbf{\bibinfo{volume}{70}}, \bibinfo{pages}{833}
  (\bibinfo{year}{1993}).

\bibitem[{\citenamefont{Ellis et~al.}(2000)\citenamefont{Ellis, Haven, and
  Turkington}}]{ellis2000}
\bibinfo{author}{\bibfnamefont{R.~S.} \bibnamefont{Ellis}},
  \bibinfo{author}{\bibfnamefont{K.}~\bibnamefont{Haven}}, \bibnamefont{and}
  \bibinfo{author}{\bibfnamefont{B.}~\bibnamefont{Turkington}},
  \bibinfo{journal}{J. Stat. Phys.} \textbf{\bibinfo{volume}{101}},
  \bibinfo{pages}{999} (\bibinfo{year}{2000}).

\bibitem[{\citenamefont{Dauxois et~al.}(2002)\citenamefont{Dauxois, Ruffo,
  Arimondo, and Wilkens}}]{dauxois2002}
\bibinfo{author}{\bibfnamefont{T.}~\bibnamefont{Dauxois}},
  \bibinfo{author}{\bibfnamefont{S.}~\bibnamefont{Ruffo}},
  \bibinfo{author}{\bibfnamefont{E.}~\bibnamefont{Arimondo}}, \bibnamefont{and}
  \bibinfo{author}{\bibfnamefont{M.}~\bibnamefont{Wilkens}}, in
  \emph{\bibinfo{booktitle}{Dynamics and Thermodynamics of Systems with Long
  Range Interactions}}, edited by
  \bibinfo{editor}{\bibfnamefont{T.}~\bibnamefont{Dauxois}},
  \bibinfo{editor}{\bibfnamefont{S.}~\bibnamefont{Ruffo}},
  \bibinfo{editor}{\bibfnamefont{E.}~\bibnamefont{Arimondo}}, \bibnamefont{and}
  \bibinfo{editor}{\bibfnamefont{M.}~\bibnamefont{Wilkens}}
  (\bibinfo{publisher}{Springer}, \bibinfo{address}{New York},
  \bibinfo{year}{2002}), vol. \bibinfo{volume}{602} of
  \emph{\bibinfo{series}{Lecture Notes in Physics}}.

\bibitem[{\citenamefont{Touchette et~al.}(2004)\citenamefont{Touchette, Ellis,
  and Turkington}}]{touchette2004}
\bibinfo{author}{\bibfnamefont{H.}~\bibnamefont{Touchette}},
  \bibinfo{author}{\bibfnamefont{R.~S.} \bibnamefont{Ellis}}, \bibnamefont{and}
  \bibinfo{author}{\bibfnamefont{B.}~\bibnamefont{Turkington}},
  \bibinfo{journal}{Physica A} \textbf{\bibinfo{volume}{340}},
  \bibinfo{pages}{138} (\bibinfo{year}{2004}).

\bibitem[{\citenamefont{Ispolatov and Cohen}(2000)}]{ispolatov2000}
\bibinfo{author}{\bibfnamefont{I.}~\bibnamefont{Ispolatov}} \bibnamefont{and}
  \bibinfo{author}{\bibfnamefont{E.~G.~D.} \bibnamefont{Cohen}},
  \bibinfo{journal}{Physica A} \textbf{\bibinfo{volume}{295}},
  \bibinfo{pages}{475} (\bibinfo{year}{2000}).

\bibitem[{\citenamefont{Barr{\'e} et~al.}(2001)\citenamefont{Barr{\'e},
  Mukamel, and Ruffo}}]{barre2001}
\bibinfo{author}{\bibfnamefont{J.}~\bibnamefont{Barr{\'e}}},
  \bibinfo{author}{\bibfnamefont{D.}~\bibnamefont{Mukamel}}, \bibnamefont{and}
  \bibinfo{author}{\bibfnamefont{S.}~\bibnamefont{Ruffo}},
  \bibinfo{journal}{Phys. Rev. Lett.} \textbf{\bibinfo{volume}{87}},
  \bibinfo{pages}{030601} (\bibinfo{year}{2001}).

\bibitem[{\citenamefont{Ellis et~al.}(2004)\citenamefont{Ellis, Touchette, and
  Turkington}}]{ellis2004}
\bibinfo{author}{\bibfnamefont{R.~S.} \bibnamefont{Ellis}},
  \bibinfo{author}{\bibfnamefont{H.}~\bibnamefont{Touchette}},
  \bibnamefont{and}
  \bibinfo{author}{\bibfnamefont{B.}~\bibnamefont{Turkington}},
  \bibinfo{journal}{Physica A} \textbf{\bibinfo{volume}{335}},
  \bibinfo{pages}{518} (\bibinfo{year}{2004}).

\bibitem[{\citenamefont{Costeniuc
  et~al.}(2005{\natexlab{a}})\citenamefont{Costeniuc, Ellis, and
  Touchette}}]{costeniuc2005}
\bibinfo{author}{\bibfnamefont{M.}~\bibnamefont{Costeniuc}},
  \bibinfo{author}{\bibfnamefont{R.~S.} \bibnamefont{Ellis}}, \bibnamefont{and}
  \bibinfo{author}{\bibfnamefont{H.}~\bibnamefont{Touchette}},
  \bibinfo{journal}{J. Math. Phys.} \textbf{\bibinfo{volume}{46}},
  \bibinfo{pages}{063301} (\bibinfo{year}{2005}{\natexlab{a}}).

\bibitem[{\citenamefont{Barr{\'e} et~al.}(2005)\citenamefont{Barr{\'e},
  Bouchet, Dauxois, and Ruffo}}]{barre2005}
\bibinfo{author}{\bibfnamefont{J.}~\bibnamefont{Barr{\'e}}},
  \bibinfo{author}{\bibfnamefont{F.}~\bibnamefont{Bouchet}},
  \bibinfo{author}{\bibfnamefont{T.}~\bibnamefont{Dauxois}}, \bibnamefont{and}
  \bibinfo{author}{\bibfnamefont{S.}~\bibnamefont{Ruffo}}, \bibinfo{journal}{J.
  Stat. Phys.} \textbf{\bibinfo{volume}{119}}, \bibinfo{pages}{677}
  (\bibinfo{year}{2005}).

\bibitem[{\citenamefont{Hahn and Kastner}(2006)}]{hahn2006}
\bibinfo{author}{\bibfnamefont{I.}~\bibnamefont{Hahn}} \bibnamefont{and}
  \bibinfo{author}{\bibfnamefont{M.}~\bibnamefont{Kastner}},
  \bibinfo{journal}{Eur. Phys. J. B} \textbf{\bibinfo{volume}{50}},
  \bibinfo{pages}{311} (\bibinfo{year}{2006}).

\bibitem[{\citenamefont{Costeniuc
  et~al.}(2005{\natexlab{b}})\citenamefont{Costeniuc, Ellis, Touchette, and
  Turkington}}]{costeniuc22005}
\bibinfo{author}{\bibfnamefont{M.}~\bibnamefont{Costeniuc}},
  \bibinfo{author}{\bibfnamefont{R.~S.} \bibnamefont{Ellis}},
  \bibinfo{author}{\bibfnamefont{H.}~\bibnamefont{Touchette}},
  \bibnamefont{and}
  \bibinfo{author}{\bibfnamefont{B.}~\bibnamefont{Turkington}},
  \bibinfo{journal}{J. Stat. Phys.} \textbf{\bibinfo{volume}{119}},
  \bibinfo{pages}{1283} (\bibinfo{year}{2005}{\natexlab{b}}).

\bibitem[{\citenamefont{Costeniuc
  et~al.}(2006{\natexlab{a}})\citenamefont{Costeniuc, Ellis, Touchette, and
  Turkington}}]{costeniuc2006}
\bibinfo{author}{\bibfnamefont{M.}~\bibnamefont{Costeniuc}},
  \bibinfo{author}{\bibfnamefont{R.~S.} \bibnamefont{Ellis}},
  \bibinfo{author}{\bibfnamefont{H.}~\bibnamefont{Touchette}},
  \bibnamefont{and}
  \bibinfo{author}{\bibfnamefont{B.}~\bibnamefont{Turkington}},
  \bibinfo{journal}{Phys. Rev. E} \textbf{\bibinfo{volume}{73}},
  \bibinfo{pages}{026105} (\bibinfo{year}{2006}{\natexlab{a}}).

\bibitem[{\citenamefont{Touchette and Beck}(2006)}]{touchette2006}
\bibinfo{author}{\bibfnamefont{H.}~\bibnamefont{Touchette}} \bibnamefont{and}
  \bibinfo{author}{\bibfnamefont{C.}~\bibnamefont{Beck}}, \bibinfo{journal}{J.
  Stat. Phys.} \textbf{\bibinfo{volume}{125}}, \bibinfo{pages}{455}
  (\bibinfo{year}{2006}).

\bibitem[{\citenamefont{Costeniuc
  et~al.}(2006{\natexlab{b}})\citenamefont{Costeniuc, Ellis, and
  Touchette}}]{costeniuc22006}
\bibinfo{author}{\bibfnamefont{M.}~\bibnamefont{Costeniuc}},
  \bibinfo{author}{\bibfnamefont{R.~S.} \bibnamefont{Ellis}}, \bibnamefont{and}
  \bibinfo{author}{\bibfnamefont{H.}~\bibnamefont{Touchette}},
  \bibinfo{journal}{Phys. Rev. E} \textbf{\bibinfo{volume}{74}},
  \bibinfo{pages}{010105} (\bibinfo{year}{2006}{\natexlab{b}}).

\bibitem[{\citenamefont{Lynden-Bell and Wood}(1968)}]{lynden1968}
\bibinfo{author}{\bibfnamefont{D.}~\bibnamefont{Lynden-Bell}} \bibnamefont{and}
  \bibinfo{author}{\bibfnamefont{R.}~\bibnamefont{Wood}},
  \bibinfo{journal}{Mon. Notic. Roy. Astron. Soc.}
  \textbf{\bibinfo{volume}{138}}, \bibinfo{pages}{495} (\bibinfo{year}{1968}).

\bibitem[{\citenamefont{Thirring}(1970)}]{thirring1970}
\bibinfo{author}{\bibfnamefont{W.}~\bibnamefont{Thirring}},
  \bibinfo{journal}{Z. Physik} \textbf{\bibinfo{volume}{235}},
  \bibinfo{pages}{339} (\bibinfo{year}{1970}).

\bibitem[{\citenamefont{Chavanis}(2002)}]{chavanis2002}
\bibinfo{author}{\bibfnamefont{P.-H.} \bibnamefont{Chavanis}},
  \bibinfo{journal}{Phys. Rev. E} \textbf{\bibinfo{volume}{65}},
  \bibinfo{pages}{056123} (\bibinfo{year}{2002}).

\bibitem[{\citenamefont{Chavanis and Ispolatov}(2002)}]{chavanis22002}
\bibinfo{author}{\bibfnamefont{P.-H.} \bibnamefont{Chavanis}} \bibnamefont{and}
  \bibinfo{author}{\bibfnamefont{I.}~\bibnamefont{Ispolatov}},
  \bibinfo{journal}{Phys. Rev. E} \textbf{\bibinfo{volume}{66}},
  \bibinfo{pages}{036109} (\bibinfo{year}{2002}).

\bibitem[{\citenamefont{Smith and O'Neil}(1990)}]{smith1990}
\bibinfo{author}{\bibfnamefont{R.~A.} \bibnamefont{Smith}} \bibnamefont{and}
  \bibinfo{author}{\bibfnamefont{T.~M.} \bibnamefont{O'Neil}},
  \bibinfo{journal}{Phys. Fluids B} \textbf{\bibinfo{volume}{2}},
  \bibinfo{pages}{2961} (\bibinfo{year}{1990}).

\bibitem[{\citenamefont{Kiessling and Lebowitz}(1997)}]{kiessling1997}
\bibinfo{author}{\bibfnamefont{M.~K.-H.} \bibnamefont{Kiessling}}
  \bibnamefont{and} \bibinfo{author}{\bibfnamefont{J.}~\bibnamefont{Lebowitz}},
  \bibinfo{journal}{Lett. Math. Phys.} \textbf{\bibinfo{volume}{42}},
  \bibinfo{pages}{43} (\bibinfo{year}{1997}).

\bibitem[{\citenamefont{Ellis et~al.}(2002)\citenamefont{Ellis, Haven, and
  Turkington}}]{ellis2002}
\bibinfo{author}{\bibfnamefont{R.~S.} \bibnamefont{Ellis}},
  \bibinfo{author}{\bibfnamefont{K.}~\bibnamefont{Haven}}, \bibnamefont{and}
  \bibinfo{author}{\bibfnamefont{B.}~\bibnamefont{Turkington}},
  \bibinfo{journal}{Nonlinearity} \textbf{\bibinfo{volume}{15}},
  \bibinfo{pages}{239} (\bibinfo{year}{2002}).

\bibitem[{\citenamefont{Lynden-Bell}(1999)}]{lynden1999}
\bibinfo{author}{\bibfnamefont{D.}~\bibnamefont{Lynden-Bell}},
  \bibinfo{journal}{Physica A} \textbf{\bibinfo{volume}{263}},
  \bibinfo{pages}{293} (\bibinfo{year}{1999}).

\bibitem[{\citenamefont{Touchette}(2003)}]{touchette22003}
\bibinfo{author}{\bibfnamefont{H.}~\bibnamefont{Touchette}}, Ph.D. thesis,
  \bibinfo{school}{McGill University} (\bibinfo{year}{2003}).

\bibitem[{\citenamefont{Hahn and Kastner}(2005)}]{hahn2005}
\bibinfo{author}{\bibfnamefont{I.}~\bibnamefont{Hahn}} \bibnamefont{and}
  \bibinfo{author}{\bibfnamefont{M.}~\bibnamefont{Kastner}},
  \bibinfo{journal}{Phys. Rev. E} \textbf{\bibinfo{volume}{72}},
  \bibinfo{pages}{056134} (\bibinfo{year}{2005}).

\bibitem[{\citenamefont{Rugh}(1998)}]{rugh1998}
\bibinfo{author}{\bibfnamefont{H.~H.} \bibnamefont{Rugh}}, \bibinfo{journal}{J.
  Phys. A: Math. Gen.} \textbf{\bibinfo{volume}{31}}, \bibinfo{pages}{7761}
  (\bibinfo{year}{1998}).

\bibitem[{\citenamefont{Rugh}(2001)}]{rugh2001}
\bibinfo{author}{\bibfnamefont{H.~H.} \bibnamefont{Rugh}},
  \bibinfo{journal}{Phys. Rev. E} \textbf{\bibinfo{volume}{64}},
  \bibinfo{pages}{055101} (\bibinfo{year}{2001}).

\bibitem[{\citenamefont{Desai and Zwanzig}(1978)}]{desai1978}
\bibinfo{author}{\bibfnamefont{R.}~\bibnamefont{Desai}} \bibnamefont{and}
  \bibinfo{author}{\bibfnamefont{R.}~\bibnamefont{Zwanzig}},
  \bibinfo{journal}{J. Stat. Phys.} \textbf{\bibinfo{volume}{19}},
  \bibinfo{pages}{1} (\bibinfo{year}{1978}).

\bibitem[{\citenamefont{Dauxois et~al.}(2003)\citenamefont{Dauxois, Lepri, and
  Ruffo}}]{dauxois2003}
\bibinfo{author}{\bibfnamefont{T.}~\bibnamefont{Dauxois}},
  \bibinfo{author}{\bibfnamefont{S.}~\bibnamefont{Lepri}}, \bibnamefont{and}
  \bibinfo{author}{\bibfnamefont{S.}~\bibnamefont{Ruffo}},
  \bibinfo{journal}{Commun. Nonlinear Sc. Num. Simul.}
  \textbf{\bibinfo{volume}{8}}, \bibinfo{pages}{375} (\bibinfo{year}{2003}).

\bibitem[{\citenamefont{Campa and Ruffo}(2006)}]{campa2006}
\bibinfo{author}{\bibfnamefont{A.}~\bibnamefont{Campa}} \bibnamefont{and}
  \bibinfo{author}{\bibfnamefont{S.}~\bibnamefont{Ruffo}},
  \bibinfo{journal}{Physica A} \textbf{\bibinfo{volume}{369}},
  \bibinfo{pages}{517} (\bibinfo{year}{2006}).

\bibitem[{\citenamefont{Ellis}(1985)}]{ellis1985}
\bibinfo{author}{\bibfnamefont{R.~S.} \bibnamefont{Ellis}},
  \emph{\bibinfo{title}{Entropy, Large Deviations, and Statistical Mechanics}}
  (\bibinfo{publisher}{Springer-Verlag}, \bibinfo{address}{New York},
  \bibinfo{year}{1985}).

\bibitem[{\citenamefont{Ellis}(1999)}]{ellis1999}
\bibinfo{author}{\bibfnamefont{R.~S.} \bibnamefont{Ellis}},
  \bibinfo{journal}{Physica D} \textbf{\bibinfo{volume}{133}},
  \bibinfo{pages}{106} (\bibinfo{year}{1999}).

\bibitem[{\citenamefont{Touchette}(2006)}]{touchette22006}
\bibinfo{author}{\bibfnamefont{H.}~\bibnamefont{Touchette}},
  \bibinfo{journal}{Physica A} \textbf{\bibinfo{volume}{359}},
  \bibinfo{pages}{375} (\bibinfo{year}{2006}).

\bibitem[{\citenamefont{Ruelle}(1969)}]{ruelle1969}
\bibinfo{author}{\bibfnamefont{D.}~\bibnamefont{Ruelle}},
  \emph{\bibinfo{title}{Statistical Mechanics: Rigorous Results}}
  (\bibinfo{publisher}{W.A. Benjamin}, \bibinfo{address}{Reading},
  \bibinfo{year}{1969}).

\bibitem[{\citenamefont{Ellis}(1995)}]{ellis1995}
\bibinfo{author}{\bibfnamefont{R.~S.} \bibnamefont{Ellis}},
  \bibinfo{journal}{Scand. Actuarial J.} \textbf{\bibinfo{volume}{1}},
  \bibinfo{pages}{97} (\bibinfo{year}{1995}).

\end{thebibliography}

\end{document}